\documentclass[twocolumn,aps,prl,preprintnumbers]{revtex4}
\usepackage{bbm}
\usepackage[latin9]{inputenc}
\usepackage{amsmath}
\usepackage{amssymb}
\usepackage{graphicx}
\usepackage{mathrsfs}
\usepackage{amsfonts}
\usepackage{amsthm}
\usepackage{color}
\usepackage{multirow}
\usepackage{txfonts}
\usepackage{lipsum}
\usepackage{gensymb}
\usepackage[colorlinks=true,citecolor=blue,linkcolor=blue,urlcolor=blue,anchorcolor=blue]{hyperref}%
\hypersetup{colorlinks=true,citecolor=blue,linkcolor=blue,urlcolor=blue}
\providecommand{\U}[1]{\protect\rule{.1in}{.1in}}
\makeatletter

\newcommand{\Rmnum}[1]{\expandafter\@slowromancap\romannumeral #1@}
\makeatother
\makeatletter
\@ifundefined{textcolor}{}
{
\definecolor{BLACK}{gray}{0}
\definecolor{WHITE}{gray}{1}
\definecolor{RED}{rgb}{1,0,0}
\definecolor{GREEN}{rgb}{0,1,0}
\definecolor{BLUE}{rgb}{0,0,1}
\definecolor{CYAN}{cmyk}{1,0,0,0}
\definecolor{MAGENTA}{cmyk}{0,1,0,0}
\definecolor{YELLOW}{cmyk}{0,0,1,0}
}
\makeatother

\begin{document}
\title{{Terahertz magnon frequency comb}}
\author{Xianglong Yao}
\author{Zhejunyu Jin}
\author{Zhenyu Wang}
\email[Present address: School of Physics and Electronics, Hunan University, Changsha 410082, China.]{}
\author{Zhaozhuo Zeng}
\author{Peng Yan}
\email[Corresponding author: ]{yan@uestc.edu.cn}
\affiliation{School of Physics and State Key Laboratory of Electronic Thin Films and Integrated Devices, University of Electronic Science and Technology of China, Chengdu 610054, China}
\begin{abstract}
{Magnon frequency comb (MFC), the spin-wave spectra composing of equidistant coherent peaks, is attracting much attention in magnonics. A terahertz (THz) MFC, combining the advantages of the THz and MFC technologies, is highly desired because it would significantly advance the MFC applications in ultrafast magnonic metrology, sensing, and communications. Here, we show that the THz MFC can be generated by nonlinear interactions between spin waves and skyrmions in antiferromagnets [Z. Jin \emph{et al}., \href{https://doi.org/10.48550/arXiv.2301.03211}{arXiv:2301.03211}]. It is found that the strength of the three-wave mixing between propagating magnons and breathing skyrmions follows a linear dependence on the driving frequency and the MFC signal can be observed over a broad driving frequency range. Our results extend the working frequency of MFC to the THz regime, which would have potential applications in ultrafast spintronic devices and promote the development of nonlinear magnonics in antiferromagnets.
}
\end{abstract}

\maketitle
\section{Introduction}\label{sec1}
The far-infrared electromagnetic spectrum in the frequency region of $0.1-30$ terahertz (THz), known as the THz band, has long been considered the last remaining scientific gap in the electromagnetic spectrum \cite{Masayoshi2007,Sirtori2002}. THz technology has been widely applied to many fields, such as the wireless communication \cite{Rappaport2019}, medical imaging \cite{Pickwell2006}, security inspection \cite{Davies2008}, etc. Photonic and/or electronic THz devices, like quantum cascade laser \cite{Li2014}, uni-travelling-carrier photodiode \cite{Ito2005}, Schottky-diode-based multipliers \cite{Maestrini2010}, and transistor-based integrated circuits \cite{John2020}, have already been realized. Very recently, the THz technology starts making its way into the field of spintronics. For example, a novel THz emitter utilizing the spin degree of freedom in magnetic materials has emerged \cite{Feng2021,Seifert2022}, which presents several unprecedented advantages, such as the ultra-broad bandwidth (up to 30 THz) \cite{Seifert2016} and flexible tunability by external magnetic fields \cite{Yang2016} or internal magnetic textures \cite{Wu2022}.

Merging THz technology with other powerful techniques can yield unique multidimensional insight into fundamental processes at ultrafast time scales. An optical frequency comb is a spectrum consisting of a sequence of discrete and equally-spaced spectral lines that represent precise marks in frequency, like an optical ruler \cite{Udem2002}. The optical frequency comb technique revolutionized optical frequency metrology and spectroscopy \cite{Kippenberg2011,Fortier2019,Picque2019} and enabled optical atomic clocks \cite{Ludlow2015,Hall2006,Hansch2006}.
A THz optical frequency comb capable of high-resolution measurement was recently demonstrated \cite{Burghoff2014,Shin2023}, which would significantly advance THz technology applications in spectroscopy, metrology, sensing, and high-speed wireless communications.
Most recently, it has been reported that a magnon frequency comb (MFC) can be generated by the nonlinear scattering between magnons and topological solitons, like skyrmion \cite{Wang2021}, vortex \cite{Wang2022} and domain wall \cite{Beining2018,Zhou2021}. However, the working frequency of MFC in ferromagnets often lies in the GHz regime, which would not be able to keep up the demand of higher-frequency operations. Extending the frequency range of the MFC to THz is therefore of fundamental interest and necessary as well for ultrafast magnonics.

Antiferromagnets with two opposite magnetic sublattices have unique advantages, such as the full freedom in magnon polarization, vanishingly small stray field, and ultrafast magnetization dynamics typically in the THz region \cite{Jungwirth2016,Baltz2018,Gomonay2018}. These properties make antiferromagnets a promising platform for magnonics \cite{Rezende2019}. Inspired by the three-wave mixing mechanism in ferromagnets, we propose to generate MFC with a THz central frequency by nonlinear interaction between breathing skyrmions and propagating magnons in antiferromagnets. By performing systematical micromagnetic simulations, we find the following differences compared with its ferromagnetic counterpart: (i) It is difficult to induce the skyrmion breathing in antiferromagnets merely by propagating magnons and an additional driving field is demanded. (ii) There is no frequency window to observe the MFC. Modeling calculations reveal an unexpectedly large critical microwave field and show a linear dependence of the three-wave coupling strength on the microwave frequency, which explain the observed two features.

The paper is organized as follows. In Sec. \ref{sec2}, we present the analytical model describing the nonlinear interaction between magnons and breathing skyrmions. The linear dispersion relation of magnons propagating in antiferromagnets is given. The dynamical equations of magnon modes involved in three-magnon processes are derived and numerically solved. Section \ref{sec3} gives the full micromagnetic simulations to verify our theoretical analysis. Conclusions are drawn in Sec. \ref{sec4}.

\section{Theoretical model}\label{sec2}
\begin{figure}
  \centering
  \includegraphics[width=0.48\textwidth]{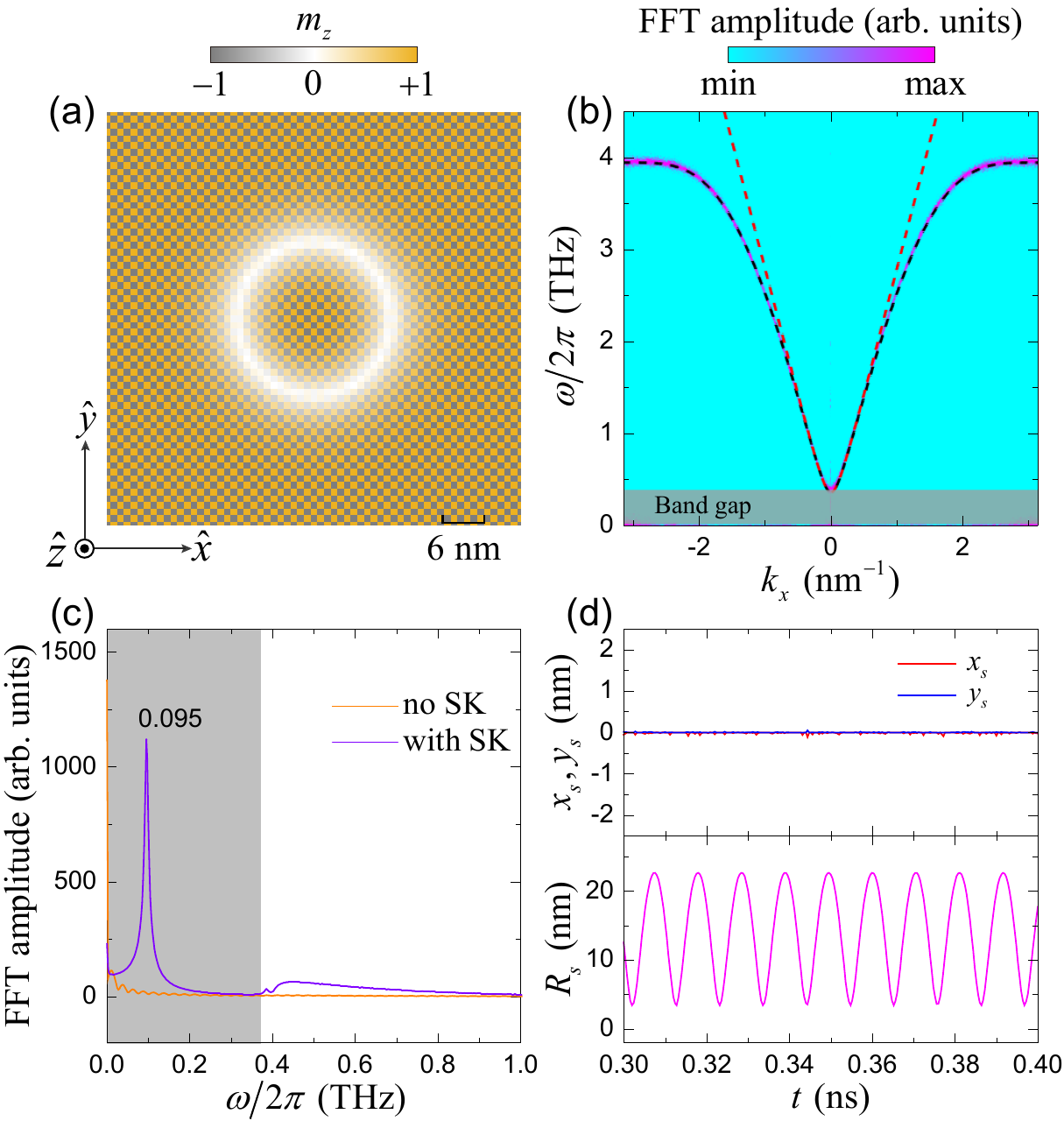}\\
  \caption{(a) Schematic diagram of an antiferromagnetic film hosting a skyrmion. (b) The dispersion relation of antiferromagnetic spin waves obtained by the FFT of the dynamical magnetization from micromagnetic simulations. The dashed black and red lines are the analytical formulas Eq. (\ref{eq_dispersion}) and Eq. (\ref{eq_dispersion2}), respectively. (c) The FFT spectra of the antiferromagnetic film with (violet curve) and without (orange curve) a skyrmion (SK). The gray regions in (b) and (c) correspond to the band gap of spin waves. (d) The time evolution of the skyrmion center position ($x_s$, $y_s$) and radius ($R_s$) under the driving field with frequency 0.095 THz.}\label{fig1}
\end{figure}

We consiser an antiferromagnetic film hosting a N$\acute{\text{e}}$el-type skyrmion stabilized by the interfacial Dzyaloshinskii-Moriya interaction (DMI), as shown in
Fig. \ref{fig1}(a). The Hamiltonian of the antiferromagnetic system can be written as
\begin{equation}\label{eq_Hamiltonian}
  \mathcal{H}=J\sum\limits_{<i,j>}\mathbf{S}_{i}\cdot\mathbf{S}_{j}-K\sum\limits_{i}(\mathbf{S}_i\cdot\hat{z})^{2}+\sum\limits_{<i,j>}\mathbf{D}_{ij}\cdot(\mathbf{S}_i\times\mathbf{S}_j),
\end{equation}
where $\mathbf{S}_i$ is the unit vector of the spin at site $i$. On the right-hand side of Eq. (\ref{eq_Hamiltonian}), the first term is the antiferromagnetic exchange interaction with coefficient $J>0$. The second term is the uniaxial magnetocrystalline anisotropy with the easy-axis along $\hat{z}$ direction and $K$ being the anisotropic constant. The third term describes the DMI where the DM vector $\mathbf{D}_{ij}=D(\hat{z}\times\mathbf{r}_{ij})$ with $D$ the DMI strength and $\mathbf{r}_{ij}$ representing the position vector connecting two neighboring spins.
Under the continuum description, the antiferromagnetic system is characterized by the net magnetization $\mathbf{m}=(\mathbf{S}_A+\mathbf{S}_B)/2$ and the staggered magnetization $\mathbf{n}=(\mathbf{S}_A-\mathbf{S}_B)/2$, where $\mathbf{S}_A$ and $\mathbf{S}_B$ are the unit magnetization of two sublattices. Then, the system Hamiltonian in Eq. (\ref{eq_Hamiltonian}) can be recast as
\begin{equation}\label{eq_Hamiltonian2}
\begin{aligned}
\mathcal{H}=&\int\bigg \{\frac{\lambda}{2}{\bf m}^{2}+\frac{A}{2}\Big [(\nabla {\bf n})
^{2}+\partial_{x}{\bf n}\cdot\partial_{y}{\bf n}\Big ]+L{\bf m}\cdot(\partial_{x}{\bf n}
\\
&+\partial_{y}{\bf n})-\frac{K_c}{2}n_{z}^{2}+\frac{D_c}{2}\Big [n_{z}\nabla\cdot {\bf n}-({\bf n}\cdot\nabla)n_{z}\Big ]\bigg \}d{\bf r},
\end{aligned}
\end{equation}
where $\lambda=8J$, $A=2Jd^2$, $L=2\sqrt{2}Jd$, $K_c=2K$, and $D_c=2Dd$ are the homogeneous exchange, inhomogeneous exchange, parity-breaking, magnetic anisotropy, and DMI constants, respectively, and $d$ is the lattice constant.

It has been demonstrated that the MFC can be generated by nonlinear coupling between magnons and breathing skyrmion in ferromagnets \cite{Wang2021}. It is naturally expected that this mechanism could also work in antiferromagnets. To investigate the nonlinear interaction between antiferromagnetic magnons and skyrmion, we express the dynamical staggered magnetization in terms of the magnon creation and annihilation operators $a^{\dag}(\textbf{r}, t)$ and $a(\textbf{r}, t)$ by using the
Holstein-Primakoff transformation. Then the Hamiltonian can be rewritten as $\mathcal{H}=\mathcal{H}^{(0)}+\mathcal{H}^{(2)}+\mathcal{H}^{(3)}+\mathcal{H}^{(4)}+\cdots$, where $\mathcal{H}^{(0)},
\mathcal{H}^{(2)},\mathcal{H}^{(3)}$, and $\mathcal{H}^{(4)}$ are the ground state energy, two-, three-, and four-magnon processes, respectively.
The dispersion relation of antiferromagnetic magnons is described by the second-order Hamiltonian $\mathcal{H}^{(2)}$, which can be expressed as
\begin{equation}\label{eq_dispersion}
  \omega=2\gamma\sqrt{(2J+K)^2-J^2[\cos(k_{x} d)+\cos(k_{y} d)]^2},
\end{equation}
with the gyromagnetic ratio $\gamma=1.76\times10^{11} \text{rad s}^{-1}\text{T}^{-1}$, and $\mathbf{k}=(k_x,k_y)$ being the wave vector of spin waves, in the lattice model. In the continue limit, the dispersion relation reads
\begin{equation}\label{eq_dispersion2}
  \omega=\gamma\sqrt{\lambda\bigg(\frac{A}{2}k^{2}+K_c\bigg)},
\end{equation}
with $k=|\mathbf{k}|$. Micromagnetic simulations agree well with the analytical formula Eq. (\ref{eq_dispersion}) [see the dashed black line in Fig. \ref{fig1}(b)], which justify the validity of the antiferromagnetic spin-wave dispersion relation. Simulation details can be found in Sec. \ref{sec3}. For Eq. (\ref{eq_dispersion2}), the numerical dispersion shows a good agreement in a low-$k$ region, but deviates in the high-$k$ region [see the dashed red line in Fig. \ref{fig1}(b)]. It is because the spin-wave dispersion relation Eq. (\ref{eq_dispersion2}) is valid only when the wavelength of spin waves is much larger the lattice distance ($2\pi/k\gg d$). According to Eq. (\ref{eq_dispersion}) [or Eq. (\ref{eq_dispersion2})], the antiferromagnetic resonance frequency is given by $\omega_{\text{AFMR}}/2\pi=(\gamma/\pi)\sqrt{4JK+K^2}=0.369$ THz, below which spin waves cannot propagate [see the gay region in Fig. \ref{fig1}(b)].

The nonlinear interaction between magnons and breathing skyrmion is described by the third-order Hamiltonian $\mathcal{H}^{(3)}$, which involves four modes: the incident magnon mode $a_k$, skyrmion breathing mode $a_r$, confluence mode $a_p$, and splitting mode $a_q$. The dynamical equations of four magnon modes can be derived as \cite{Slobodianiuk2019}
\begin{equation}\label{sw_dynamics}
\begin{aligned}
&i\frac{da_{k}}{dt}=(\omega_{k}-i\Gamma_{k})a_{k}+g_{q}a_{r}a_{q}+g_{p}a_{r}^{\dag}a_{p}+h_{1}e^{-i\omega_{1}t},\\
&i\frac{da_{r}}{dt}=(\omega_{r}-i\Gamma_{r})a_{r}+g_{q}a_{k}a_{q}^{\dag}+g_{p}a_{k}^{\dag}a_{p}+h_{2}e^{-i\omega_{2}t},\\
&i\frac{da_{p}}{dt}=(\omega_{p}-i\Gamma_{p})a_{p}+g_{p}a_{k}a_{r},\\
&i\frac{da_{q}}{dt}=(\omega_{q}-i\Gamma_{q})a_{q}+g_{q}a_{k}a_{r}^{\dag},
\end{aligned}
\end{equation}
where $\Gamma_{v}=\alpha_{v}\omega_{v}\ (v=k,r,p,q)$ are the damping rate of the magnon modes with the effective Gilbert damping constants $\alpha_{v}$ and mode frequencies $\omega_{v}$. $g_{p}$ and $g_{q}$ are the coupling strength of the three-magnon confluence and splitting, respectively. The incident spin wave is excited by the microwave field $h_{1}e^{-i\omega_{1}t}$. Since the skyrmion breathing mode in antiferromagnets is difficult to be excited merely by the incident spin waves in a non-resonant manner, as shall be discussed later, we apply another microwave field $h_{2}e^{-i\omega_{2}t}$ to resonantly excite the skyrmion breathing mode. Here, $h_i$ and $\omega_i$ ($i=1,2$) are the microwave field amplitude and frequency, respectively. Then, the incident spin-wave mode ($\omega_{k}$) and breathing mode ($\omega_{r}$) of the skyrmion would mix with each other and generate the sum-frequency ($\omega_{p}=\omega_{k}+\omega_{r}$) and difference-frequency ($\omega_{q}=\omega_{k}-\omega_{r}$) modes. These two secondary signals further hybridize with the skyrmion breathing mode to generate higher-order frequency modes, eventually leading to the MFC.

\section{Micromagnetic simulations}\label{sec3}

To verify the above picture, we perform full micromagnetic simulations using MUMAX3 \cite{Vansteenkiste2014}. We consider a G-type antiferromagnetic film with dimensions $1000\times1000\times1$ nm$^{3}$. The cell size of $1\times1\times1$ nm$^{3}$ is used to discrete the film in simulations. Magnetic parameters of KMnF$_{3}$ are adopted \cite{Barker2016,Shen2020}: $M_s=3.76\times10^{5}$ A/m, $A^{\mathrm{sim}}=A M_s/4=6.59$ pJ/m, $K^{\mathrm{sim}}=K_c M_s/2=1.16\times10^{5}$ J/m$^{3}$, $D^{\mathrm{sim}}=D_c M_s/2=1$ mJ/m$^{2}$, and $\alpha=1\times10^{-3}$. Absorbing boundary conditions are used to avoid the spin-wave reflection by film edges.

To characterize the intrinsic modes of antiferromagnetic skyrmion, we apply a sinc-function field $\mathbf{h}(t)=h_{0}\mathrm{sinc}(\omega_{H}t)\hat{z}$ with amplitude $h_0=10$ mT and cutoff frequency $\omega_H/2\pi=2.5$ THz over the antiferromagnetic film. By carrying a standard fast Fourier transformation (FFT) for each cell and then averaging over the whole film, we find one main peak at 0.095 THz in the band gap [see the violet curve in Fig. \ref{fig1}(c)]. For the film without a skyrmion, this peak is absent [see the orange curve in Fig. \ref{fig1}(c)].
This indicates that the peak at 0.095 THz comes from the intrinsic mode of the antiferromagnetic skyrmion. By analyzing the skyrmion motion under a sinusoidal microwave field with frequency 0.095 THz, we identify this mode as the breathing mode ($\omega_r$) of the antiferromagnetic skyrmion, as shown in Fig. \ref{fig1}(d). Here, the position and radius of skyrmion are obtained by a circular curve fitting of the $m_z=0$ contour.


\begin{figure}
  \centering
  \includegraphics[width=0.48\textwidth]{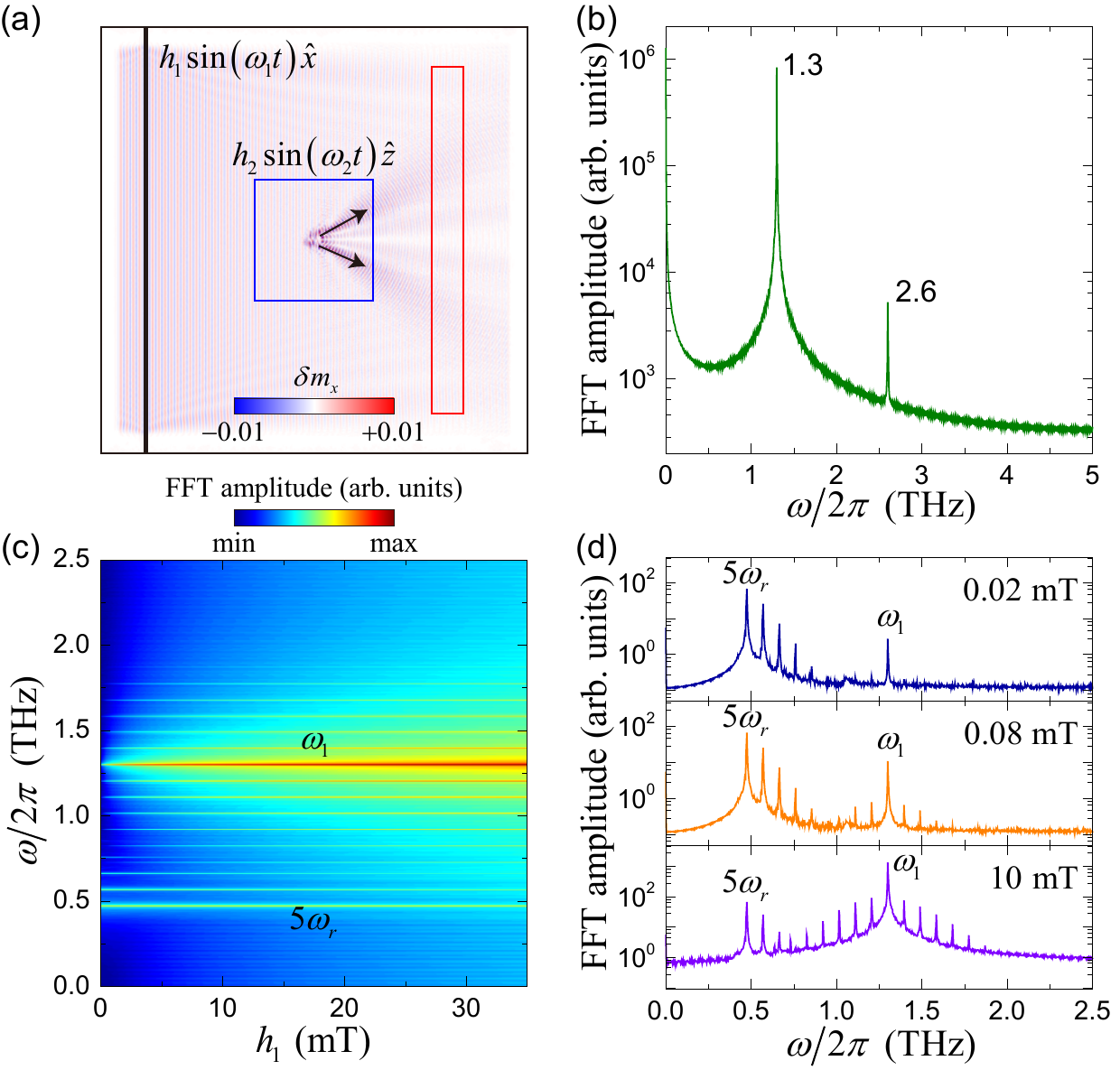}\\
  \caption{(a) Snapshot of the interaction between propagating magnons and skyrmion in AFM. The incident spin waves and breathing skyrmion are excited by the microwave fields $h_1\sin(\omega_1 t)\hat{x}$ (black bar) and $h_2\sin(\omega_2 t)\hat{z}$ (blue box), respectively. The red rectangle is the detection region. Black arrows label the Hall trajectory of magnons. (b) The spin-wave spectrum of the detection region in (a) when only the exciting field $h_1=500$ mT is applied. (c) The spin-wave spectra as a function of the driving field amplitude ($h_1$). The driving frequency is fixed at 1.3 THz. (d) The FFT spectrum at three representative fields $h_1=0.02$ mT (upper panel), 0.08 mT (middle panel), and 10 mT (lower panel), respectively. In (c) and (d), the skyrmion breathing mode is excited by the second microwave field with amplitude $h_2=5$ mT and frequency $\omega_2/2\pi=0.095$ THz.}\label{fig2}
\end{figure}

In ferromagnets, when the amplitude of the local exciting field increases above a threshold, the skyrmion breathing would be excited via a three-wave splitting process. Subsequent wave-mixing would result in the MFC \cite{Wang2021}. Inspired by this idea, we first apply a linearly-polarized microwave field $\mathbf{h}(t)=h_1\sin(\omega_1 t)\hat{x}$ with $\omega_{1}/2\pi=1.3$ THz in a narrow rectangular area [black bar in Fig. \ref{fig2}(a)] to excite the incident spin waves, which then interact with the skyrmion for generating the MFC. However, there is no sign of the MFC except the incident spin-wave mode and its frequency-doubling signal, even when the field amplitude increases up to 500 mT, as shown in Fig. \ref{fig2}(b), which is already beyond the scope of conventional experiments. It is noted that the skyrmion breathing in chiral antiferromagnets can be conveniently excited by a modification of the DMI or magnetocrystalline anisotropy \cite{Qiu2021,Komineas2022}.

To overcome this obstacle, we apply another microwave field $\mathbf{h}_2=h_2\sin(\omega_2 t)\hat{z}$ with $\omega_2/2\pi=0.095$ THz ($\omega_2=\omega_r$) in a square region [blue box in Fig. \ref{fig2}(a)] to excite the skyrmion breathing mode. Then, the incident magnons would interact with the breathing skyrmion. Black arrows represent the trajectories of scattered magnons, which may include both the linear and nonlinear topological magnon spin Hall effects \cite{Jin2023}. The experimental scheme for detection shall be discussed below. To determine whether the MFC is generated or not, we detect the magnon spectrum in a rectangular region behind the skyrmion [red rectangle in Fig. \ref{fig2}(a)] by performing FFT of the dynamical magnetization. Figure \ref{fig2}(c) shows the FFT spectra by continuously varying the microwave field amplitude $h_1$. Around 0.5 and 1.3 THz, two sets of magnon signals can be observed. The modes around 0.5 THz correspond to the frequency multiplication of the skyrmion breathing mode of frequency $n\omega_r$ with $n$ being an integer. Because of the band gap limited by the antiferromagnetic resonance $\omega_{\text{AFMR}}$, only the high-order modes above 0.369 THz ($n\geq5$) can escape from the skyrmion and are subsequently detected in the red rectangle region. The modes around 1.3 THz are the MFC modes generated by nonlinear interaction between the incident magnons and skyrmion breathing mode. Since the skyrmion breathing mode is directly excited by the microwave field $\mathbf{h}_2$ rather than the incident magnon, the driving field amplitude is not required to be larger than the aforementioned threshold value ($\gg500$ mT). Nevertheless, when the incident spin waves are excited by a weak driving field ($h_1=0.02$ mT), the MFC signals can hardly be observed [see the upper panel in Fig. \ref{fig2}(d)]. This is mainly because the amplitudes of newly generated MFC modes are very small and decay rapidly during their propagation in the presence of Gilbert damping. By increasing the driving field amplitude, the MFC modes with frequency spacing equal to $\omega_r$ clearly emerge, as shown in the middle and lower panels of Fig. \ref{fig2}(d).

\begin{figure}
  \centering
  \includegraphics[width=0.48\textwidth]{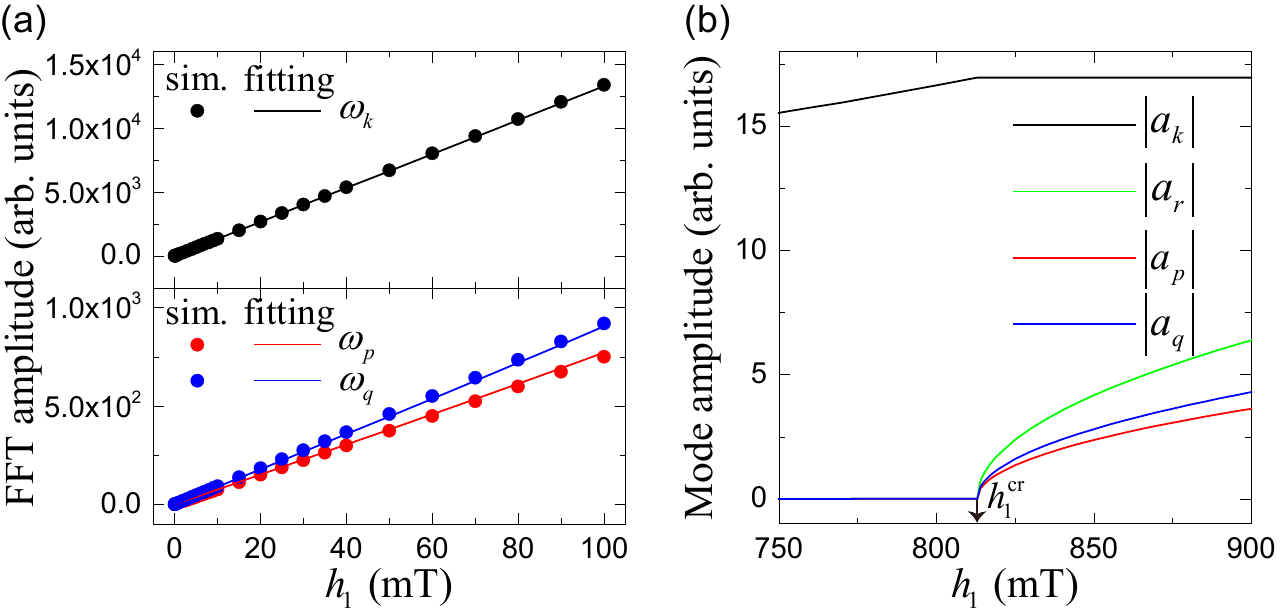}\\
  \caption{(a) The amplitudes of three main peaks at $\omega_k=\omega_1$ and $\omega_k\pm\omega_r$ as a function of the driving field amplitude ($h_1$). The solid lines are the fitting curves with the parameter $g=55.7$ MHz based on Eq. (\ref{sw_dynamics}). (b) Analytical curves of the mode amplitudes, which are obtained by numerically solving Eq. (\ref{sw_dynamics}) with parameters $\omega_k/2\pi=1.3$ THz, $\omega_r/2\pi=\omega_2/2\pi=0.095$ THz, $h_2=0$ mT, $\alpha=0.001$, and $g=55.7$ MHz.}\label{fig3}
\end{figure}

The coupling strength ($g_{p,q}$) between spin waves and skyrmion is crucial for the formation of MFC, which depends on the mode overlap and is difficult to calculate in our case due to the lacking of the analytical skyrmion profile. Hence, we treat them as fitting parameters. For simplicity, we assume $g_{p,q}=g$ and that all modes have the same damping rates $\alpha_{k,r,p,q}=\alpha$. By numerically solving Eq. (\ref{sw_dynamics}) and fitting the amplitudes of the incident, confluence, and splitting magnon modes at steady state (see Fig. \ref{fig6} in Appendix), we obtain the coupling strength $g=55.7$ MHz [see Fig. \ref{fig3}(a)]. Substituting this coupling strength back into Eq. (\ref{sw_dynamics}), we can calculate the threshold field for the excitation of the skyrmion breathing mode by propagating magnons. We identify the critical value to be $h_{1}^{\text{cr}}=813$ mT, as shown in Fig. \ref{fig3}(b). It is too high compared to the case in ferromagnets. This result provides an explanation why propagating magnons alone is not able to split the skyrmion breathing, as noticed in our micromagnetic simulations.

\begin{figure}
  \centering
  \includegraphics[width=0.48\textwidth]{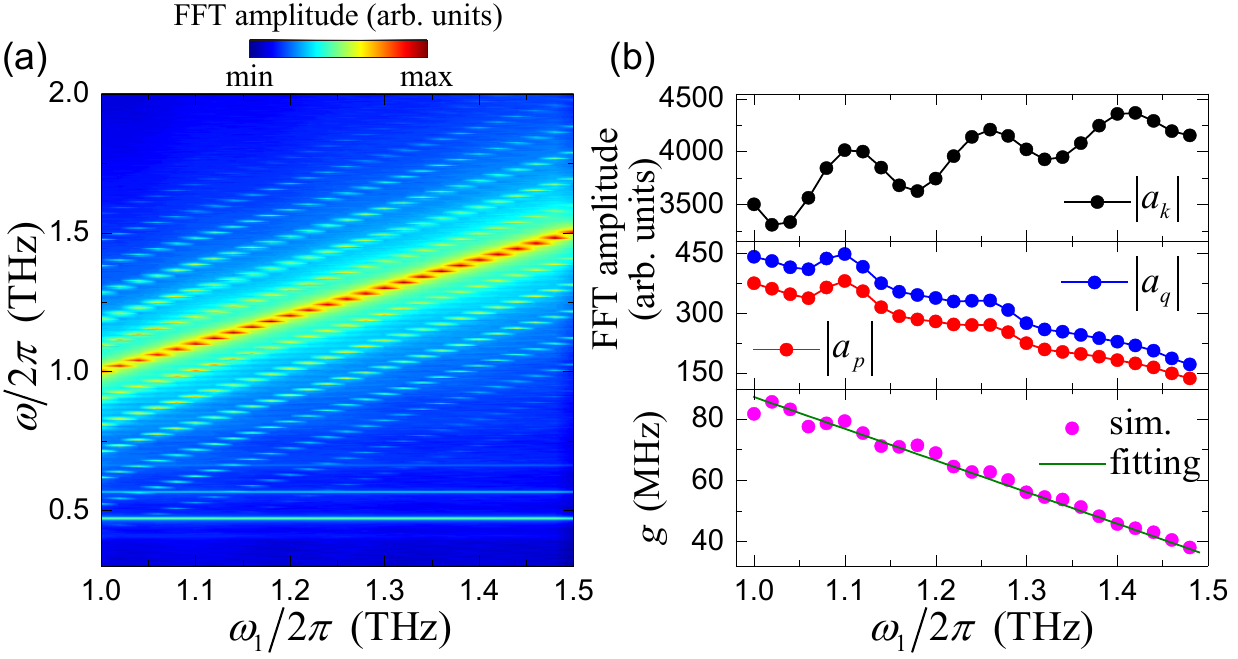}\\
  \caption{(a) The spin-wave spectrum as a function of the driving field frequency $\omega_1$. The driving field amplitude is fixed at $h_1=30$ mT. (b) The FFT amplitudes of three modes at frequencies $\omega_1$ (upper panel), and $\omega_1\pm\omega_r$ (middle panel) as a function of the driving field frequency. The lower panel in (b) shows the dependence of the coupling strength on the driving field frequency. Magenta dots are extracted from simulation results and the green line is the linear fitting.}\label{fig4}
\end{figure}

Next, we investigate the dependence of the MFC on the driving frequency, as shown Fig. \ref{fig4}(a). We find that, over a rather broad driving frequency range, the MFC is still visible and the mode spacing of the MFC is always equal to the skyrmion breathing frequency. This result significantly differs from the MFC generated in ferromagnets, where the MFC can only be observed in a frequency window and we attributed it to the Gaussian profile of the frequency-dependent three-wave coupling \cite{Wang2021}. By the same fitting method used in Fig. \ref{fig3}(a), we extract the coupling strengths for different driving frequencies, as shown in Fig. \ref{fig4}(b). One can see that the coupling strength $g$ monotonically decreases with the increase of the driving frequency $\omega_1$, approximately following a linear function $g=c\omega_1+g_0$ with the dimensionless parameter $c=-1.04\times10^{-4}$ and intercepting coupling $g_0=0.191$ THz. This result suggests that the nonlinear three-wave coupling between magnons and breathing skyrmion in ferromagnets and antiferromagnets has very different frequency dependencies, which might originate from distinct internal modes of the skyrmion in ferromagnets and antiferromagnets \cite{Kravchuk2019}. It might also be because the dispersion of magnons is parabolic in ferromagnets, but linear in antiferromagnets. Further investigations are necessary to elucidate the underlying physical origin behind such interesting findings, which goes beyond the scope of this paper.

\begin{figure}
  \centering
  \includegraphics[width=0.48\textwidth]{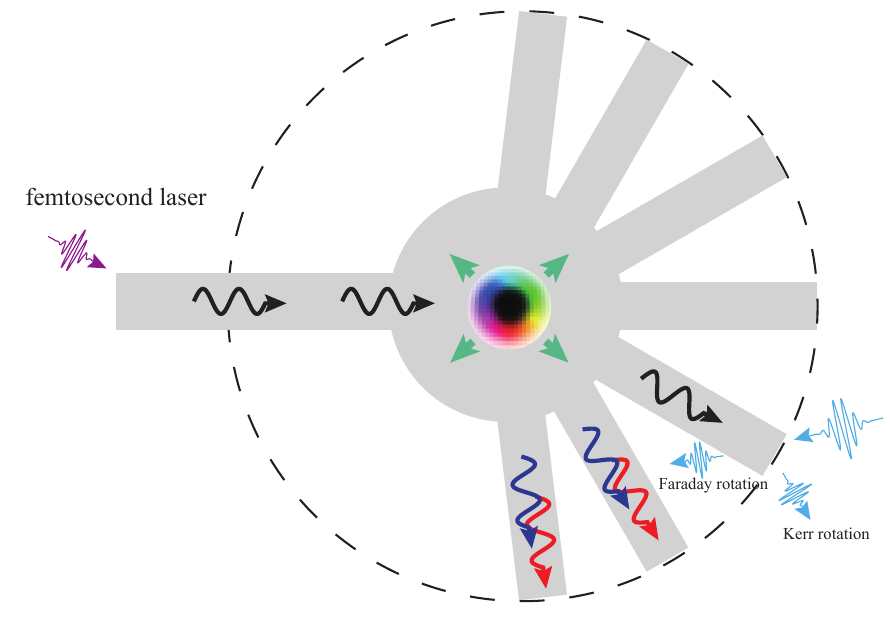}\\
  \caption{Schematic of experimental detection of nonlinear topological magnon spin Hall effect. THz magnons are excited by femtosecond laser (purple arrow) and scattered by breathing skyrmion (green arrows). Meanwhile, incident (black arrows) and nonlinear magnons (blue/red arrows) are scattered to different channels (gray background) due to the very different Hall angles. At the end of each channel, the accumulated magnons with different frequencies are detected through magneto-optical spectroscopy (blue arrows). Here, the Kerr rotation and Faraday rotation describe the rotation of the plane of polarization for reflected and transmitted laser, respectively. It is noted that left-handed magnons will be scattered to top right channels (not shown). }\label{fig5}
\end{figure}

In Ref. \cite{Jin2023}, we have identified a so-called nonlinear topological magnon spin Hall effect by analyzing the real-space scattering patterns of the MFC. Below, we propose a multi-channel device to measure it, by minimizing the spatial overlap between different magnon modes, as shown in Fig. \ref{fig5}. Experimentally, we can excite THz magnons in antiferromagnets by femtosecond laser \cite{Hortensius2021}. When the generated magnons interact with the breathing skyrmion, the MFC emerges and the incident and nonlinear magnons are scattered to different channels due to their very different Hall angles. One can distinguish the accumulated magnons with different frequencies by magneto-optical effects (e.g., Faraday or Kerr rotations), as plotted in Fig. \ref{fig5}.

\section{Conclusion}\label{sec4}
In summary, we theoretically demonstrated that the THz MFC can be generated in antiferromagnetic film by nonlinear interactions between magnons and skyrmions. Although the mechanism of the MFC generation in both ferromagnets and antiferromagnets comes from the three-wave mixing, there exist significant differences between them. First, the skyrmion breathing in antiferromagnets can hardly be excited merely by propagating magnons because of an unexpectedly large threshold field amplitude of microwaves. An additional low-frequency driving source is therefore needed to assist the MFC generation. Second, the dependence of the coupling strength on the driving frequency in two magnetic systems is different. The frequency dependence follows a Gaussian profile in ferromagnets \cite{Wang2021}, while it exhibits a linear dependence in antiferromagnets. As a consequence, the MFC in antiferromagnets is visible over a broad driving frequency range, in contrast to a narrow window in ferromagnets. From the application point of view, the THz MFC we predicted can be utilized to detect magnetic textures or defects in antiferromagnets, which is difficult to realize by conventional means because of the vanishingly small net magnetization. Our findings bring the MFC to the THz regime, which would advance MFC technology applications in ultrafast magnonic metrology, sensing, and communications.

\section{Acknowledgments}
\begin{acknowledgments}
We thank H. Yang, L. Song, and X. Liu for helpful discussions.
This work was funded by the National Key R$\&$D Program under Contract No. 2022YFA1402802 and the National Natural Science Foundation of China (NSFC) (Grants No. 12374103 and No. 12074057). Z.W. acknowledges financial support from the NSFC under Grant No. 12204089. X.Y., Z.J., and Z.W. contributed equally.
\end{acknowledgments}

\section*{APPENDIX}

\begin{figure}[htbp!]
  \centering
  \includegraphics[width=0.3\textwidth]{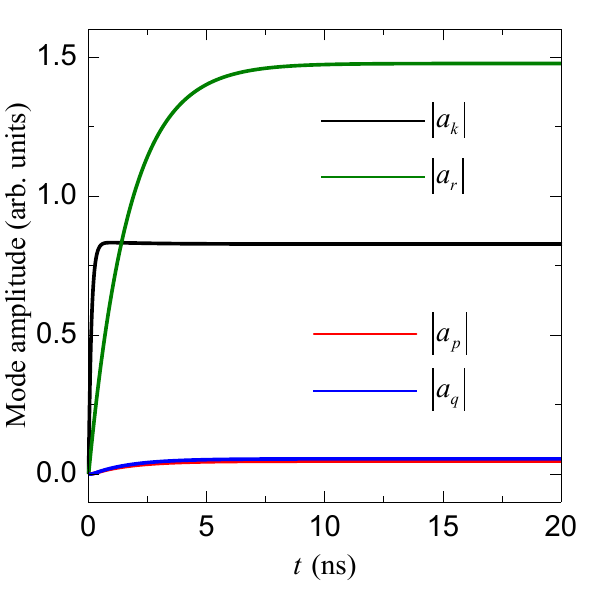}\\
  \caption{The time-evolution of four magnon modes obtained from numerically solving Eq. (\ref{sw_dynamics}). The exciting field of the incident magnon and skyrmion breathing modes are set as $h_1=40$ mT, $\omega_1/2\pi$=$\omega_k/2\pi=1.3$ THz, $h_2=5$ mT, and $\omega_2/2\pi$=$\omega_r/2\pi=0.095$ THz. The coupling strength is $g=55.7$ MHz and the damping constant is $\alpha=0.001$. }\label{fig6}
\end{figure}

In general, the dynamical equations of magnon modes Eq. (\ref{sw_dynamics}) is hard to solve analytically. Thus, we numerically solve it by using the ode45 solver in MATLAB. The initial amplitude of four modes are set as $a_k=0$, $a_r=0.001$, $a_p=0$, and $a_q=0$. After a long-time evolution (20 ns), a stationary solution can be obtained, as shown in Fig. \ref{fig6}.

\end{document}